# Giant Seebeck Effect in PEDOT Materials with Molecular Strain

*Yuki OSADA, Ryo TAKAGI, Hideki ARIMATSU, Takuya FUJIMA\**

1-28-1 Tamazutsumi Setagaya, 158-8557 Tokyo, Japan

E-mail : tfujima@tcu.ac.jp

**Abstract**

Poly 3,4-ethylenedioxythiophene (PEDOT) has been attracting attention as a thermoelectric material for room-temperature use due to its flexibility and non-toxicity. However, PEDOT reportedly generates insufficient thermoelectric power for practical use. This work tried to improve the Seebeck coefficient by introducing molecular strain to PEDOT molecules by loading a Polystyrene sulfonate (PSS)-free PEDOT on a Polyethyleneterephthalate (PET) fiber. Raman spectroscopy revealed the PEDOT materials with significant compression in the $C\alpha$-$C\alpha$ bond and extension in the $C\alpha$=$C\beta$ bond exhibit Seebeck coefficients two orders of magnitude larger than usual. Furthermore, strain in the $C\beta$-$C\beta$ bond strongly correlated with the Seebeck coefficient that varied in a broad range from -2100 to 3100 $\mu V\ K^{-1}$. This variation indicated that the molecular strain formed a sharp peak or valley around the Fermi level in the density of state (DOS) function, which gradually shifts along with the $C\beta$-$C\beta$ strain. This molecular strain-induced giant Seebeck effect is expected to be an applicable technique for other polythiophene molecules.

## 1. Introduction

In recent years, the spread of the Internet of things (IoT) devices has created an increasingly strong need for independent power sources that do not rely on large-scale centralized power sources from the perspective of ease of installation and continuity of operation.[1, 2, 3, 4]

Thermoelectric generation is expected to be an energy harvesting technology that converts waste heat into electricity, and bismuth tellurium alloys have been put to practical use in the room temperature range.[5, 6, 7, 8] In particular, organic thermoelectric materials are attracting attention as promising room-temperature thermoelectric materials due to their flexibility, light weight, and low toxicity.[9, 10, 11, 12] However, these materials are still in the research stage.

Some organic compounds exhibit the giant Seebeck effect. The mechanism of generating this large thermoelectric power has not yet been elucidated. However, it has been discussed that it is due to the interaction between phonons and carriers.[13, 14, 15, 16, 17]

The PEDOT/PSS composite material is a conductive polymer material that has been widely studied due to its high conductivity and chemical stability.[18, 19] We have also studied PEDOT conductive films with high conductivity by using hierarchical nanoporous layer (HNL) glass to eliminate the need for PSS, an insulating film stabilizer, and by introducing a macro-separated structure of PEDOT and PSS.[20, 21] The PEDOT has also been studied as a thermoelectric material, but its Seebeck coefficient is only a few tens of $\mu$V K$^{-1}$; increasing its output power is still an issue.[22, 23, 24]

In this study, we attempted to improve the thermoelectric performance of PEDOT, especially the Seebeck coefficient, by changing its molecular conformation, which is done by loading PSS-free PEDOT on the fiber material's surface.

Kirihara *et al.* carried out a similar investigation by attaching a commercially available PEDOT:PSS on a felt fabric to find out its Seebeck coefficient still comparable to that on flat substrate.[25] Therefore, we polymerized PEDOT by ourselves.

## 2. Experimental

For our study, we used PSS-free PEDOT with PTSA as a dopant supported on fibers. Flat glass slides were also used as substrates to study the effect of substrate. For comparison with the widely studied PEDOT:PSS, commercially available PEDOT:PSS was coated on each substrate.

PEDOT:PTSA was supported on the substrate (felt fabric [100% PET] and soda-lime glass) in an in-situ manner by immersing it in a PEDOT polymerization reaction solution.

For the polymerization reaction solution, 12 [$\mu$mol L$^{-1}$] of 3, 4-Ethylenedioxytiophene (EDOT >98.0%, Tokyo Chemical Industry Co., Ltd.), 14 [mmol

L$^{-1}$] of Sodiumperoxyodisulfate (SPS >97.0%, Fujifilm Wako Pure Chemicals Co., Ltd.), and 1.9 [mmol L$^{-1}$] of p-toluenesulfonic acid (PTSA >99.0%, Fujifilm Wako Pure Chemicals Co., Ltd.) were mixed and dispersed in purified water. The samples were kept in the solution for 24 hours and then dried at 336 K.

Samples with commercially available PEDOT:PSS (high-conductivity grade 0.5–1wt%, Sigma Aldrich) were also prepared for comparison. We made the felt fabric support the PEDOT:PSS through nine repetitions of immersion in the purchased PEDOT:PSS dispersion then dried it at 336 K. The dispersion was spin-coated onto the soda-lime glass.

We observed each sample through a scanning electron microscope (JCM-6000PLUS Neo scope, JEOL), Raman scattering spectroscopy (TS6400, HORIBA) using an argon laser (514.5 nm), sheet resistivity evaluation through the four-terminal method using LCR meter (ZM2376, NF Corp.), and thermoelectric power measurement.

We used a self-made system for the thermoelectric power measurement. Peltier devices controlled the high- and low-temperature sides individually to apply a temperature difference of 2–10 K while keeping the average temperature at 303 K. For the felt fabric samples, the temperature difference was applied in the normal direction of the surface for temperature stability and the in-plane direction for the glass substrate samples because the conductive film was formed only on one side.

## 3. Results and Discussion

When glass and felt fabrics were used as substrates, the samples turned dark blue after the preparation process, as shown in **Figure 1**, indicating that PEDOT was present. Since PEDOT is a hydrophobic molecule, it adheres well to PET, which is hydrophobic, but not to glass, which is hydrophilic, resulting in macroscopically non-uniform adhesion of the PEDOT:PTSA/Glass sample. The commercial PEDOT:PSS sample adhered well to glass because it contained the hydrophilic, film-forming agent PSS, but it was difficult to adhere to felt fabric, requiring nine repetitions of the adhesion process described above.

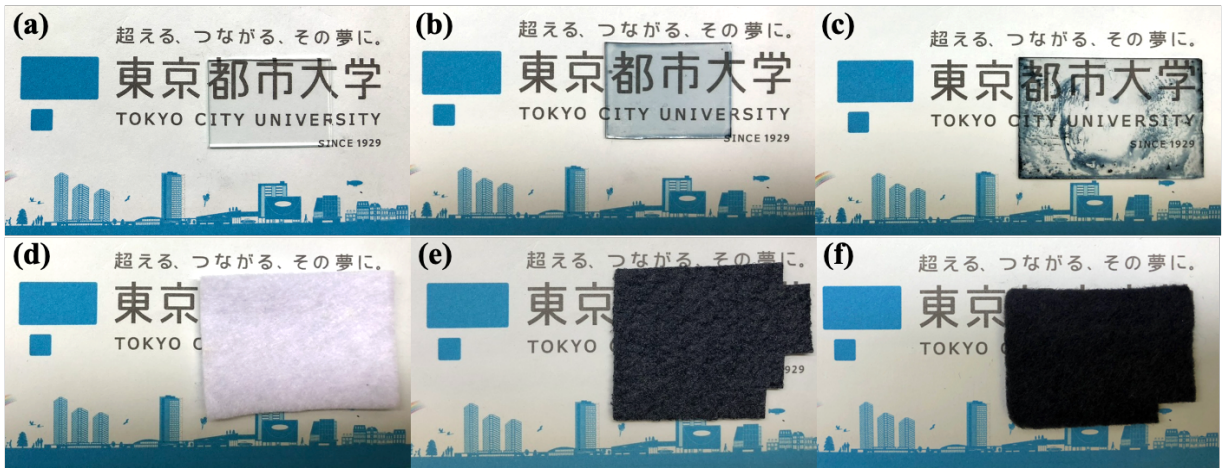

**Figure 1.** The appearance of the samples (a,d) without any treatment, (b,e) with commercial PEDOT:PSS coated, and (c,f) with PEDOT:PTSA coated. Two substrates, (a,b,c) glass slide and (d,e,f) felt fabric, were used.

**Figure 2** shows the SEM micrographs of the samples with felt fabric substrate before and after the PEDOT loading treatment. Before the treatment, the image was easily distorted by the charge due to the lack of conductivity, but for both PEDOT:PTSA and commercial PEDOT:PSS samples, the image was clearer, indicating their conductivity. Both PEDOT materials are evenly adhered to along the fibers.

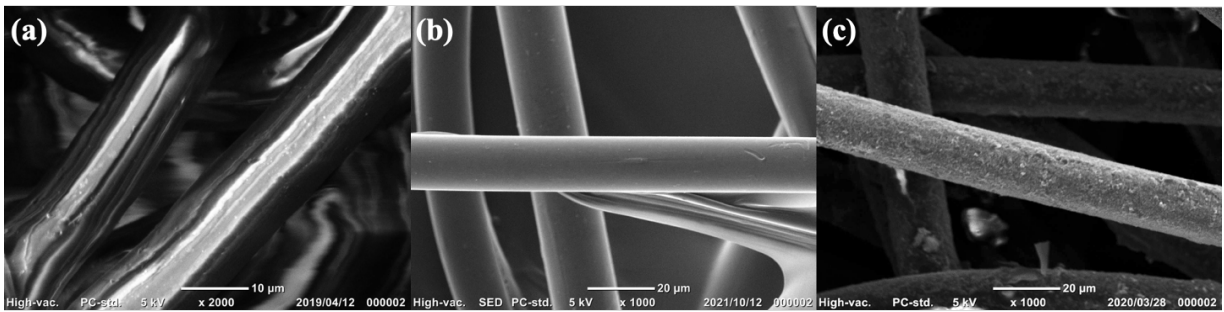

**Figure 2.** SEM micrographs of the felt fiber (a) without any treatment, (b) with PEDOT:PSS, and (c) with PEDOT:PTSA.

As shown in **Figure 3**, the sheet resistivity of all the samples decreased, indicating conductivity. Only PEDOT:PTSA/Glass showed high resistivity, probably because the hydrophobic PEDOT did not adhere easily to the hydrophilic glass surface. The other samples showed low resistivity comparable to each other, indicating that the loading and doping of PEDOT on the felt fabric was carried out comparably to those of the commercial PEDOT:PSS of high conductivity grade.

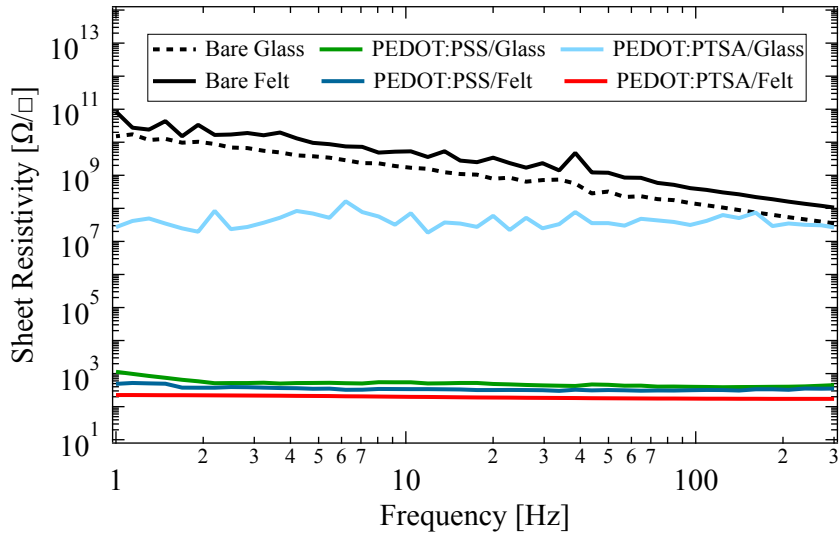

**Figure 3.** The sheet resistivity spectra of the samples.

**Figure 4** shows the temperature difference dependence of thermoelectric power. The PEDOT:PTSA/Glass with the large resistivity and samples using commercial PEDOT:PSS showed values of about ten μV K$^{-1}$, comparable to those of the general PEDOT materials reported so far. On the other hand, PEDOT:PTSA/Felt generated about two orders of magnitude larger thermoelectric power, -2100 μV K$^{-1}$, than the commercial PEDOT:PSS, despite its high conductivity comparable to them.

The Raman spectra of each sample are shown in **Figure 5**. The obtained spectra were subjected to curve fitting analysis using the peaks detected in PEDOT and the supporting materials, felt or glass, as Gaussian functions. The baseline was assumed to be a straight line. As shown in the figure, the fitting analysis reproduced the experimental data well.

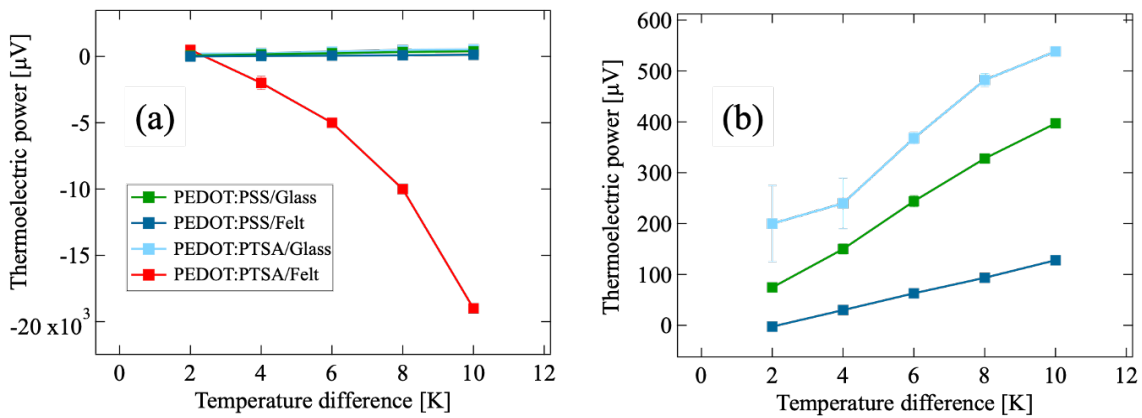

**Figure 4.** The thermoelectric power of (a) the PEDOT-coated samples and (b) the samples except for PEDOT:PTSA on felt fabric.

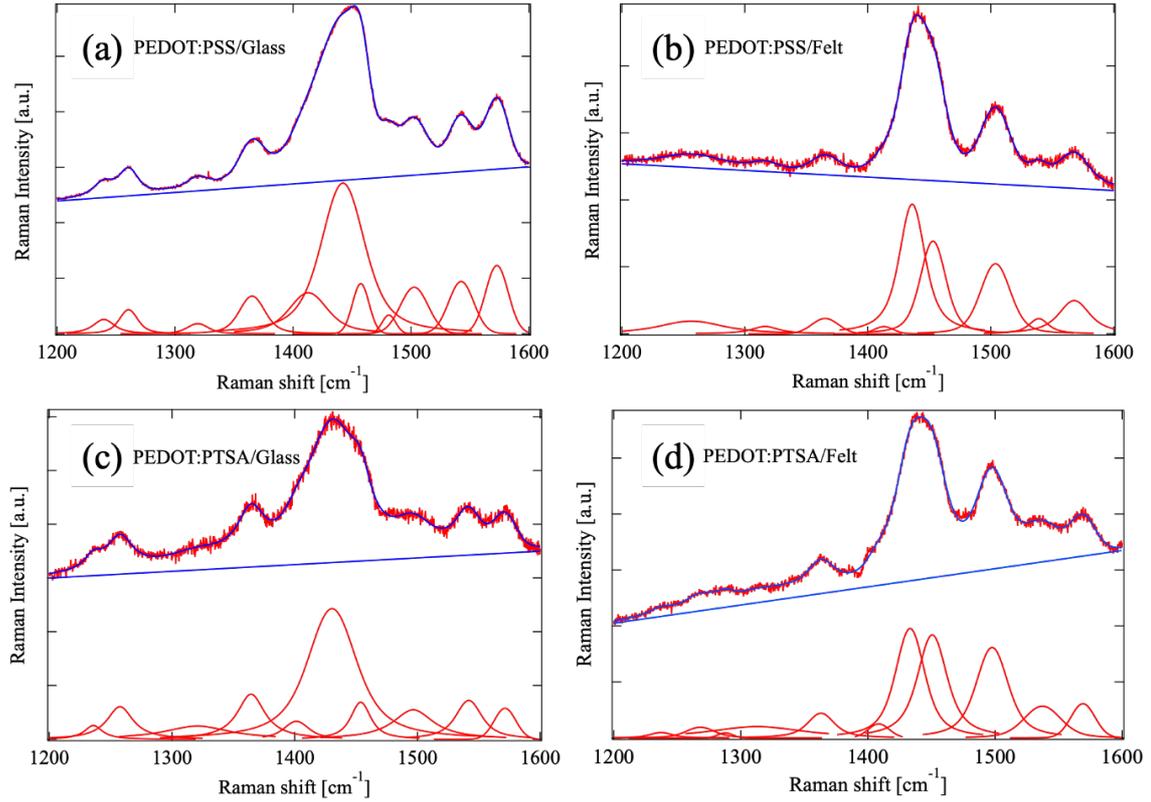

**Figure 5.** Raman spectra of (a) PEDOT:PSS on glass, (b) PEDOT:PSS on felt, (c) PEDOT:PTSA on glass, (d) PEDOT:PTSA on felt. Each spectrum (red line with noise) was reproduced (blue line along with the spectrum) with a straight baseline (blue) and Gaussian peaks corresponding to chemical bonds in samples (red smooth line).

**Figure 6(a–c)** shows the details of the three interatomic bonds that serve as conduction paths among the scattering peaks separated by the fitting analysis. It shows that the Cα-Cα bond shifts to the higher wavenumber side and that the Cα=Cβ bond only shifts to the lower wavenumber side for the sample that shows a large Seebeck coefficient. On the other hand, the Cβ-Cβ coupling does not show such a pronounced shift as the other two.

**Figure 6(d)** shows the correlation between the Raman shifts of the Cα-Cα and Cα=Cβ bonds in the PEDOT chains of each sample. This figure shows that only the felt-supported PEDOT/PTSA exhibits a molecular strain that is significantly different from that of commercial PEDOT/PSS and flat-supported PEDOT/PTSA.

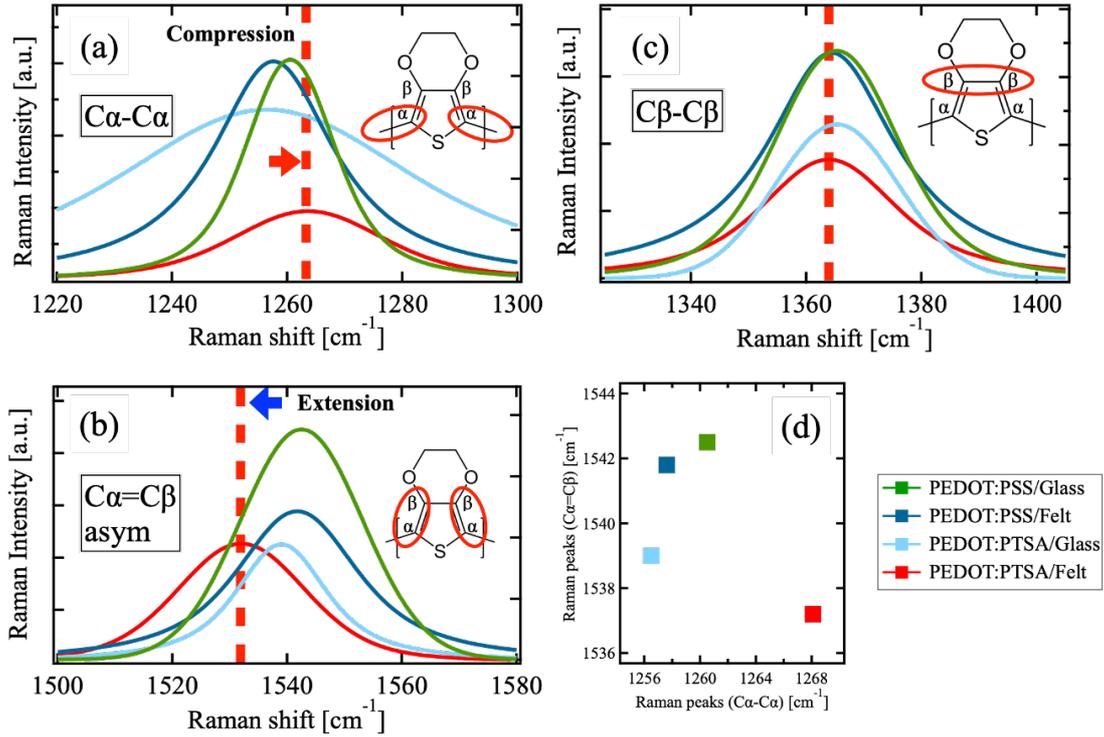

**Figure 6.** Raman scattering components for (a) Cα-Cα bond, (b) Cα=Cβ asymmetric stretching, (c) Cβ-Cβ bond in PEDOT chain, and (d) correlation between the Cα-Cα and the Cα=Cβ bonds.

There was no significant change in the Raman shift corresponding to the Cβ-Cβ bond, but there was a strong correlation (correlation coefficient of 0.91) with the Seebeck coefficient. **Figure 7** shows the correlations in the felt-supported PEDOT/PTSA prepared with the PTSA concentration varying during polymerization. In these samples, the Raman scattering peak area ratios of the benzoidal and quinoid Cα=Cβ symmetric vibration were almost identical, indicating that the doping level of the PEDOT is comparable. Therefore, this change in the Seebeck coefficient is not considered to be due to the difference in carrier concentration.

These results suggest that the giant Seebeck coefficient obtained for felt-supported PEDOT/PTSA is due to the change in the quantum state of the carriers caused by the molecular distortion. In particular, the fact that the Seebeck coefficient changes significantly between positive and negative with the Raman shift of the Cβ-Cβ bond suggests that the density of states (DOS) near the Fermi level, which determines the Seebeck coefficient, has a steep peak or valley that shifts to the valence or conduction band as the Cβ-Cβ bond is stretched or compressed. For this reason, the Seebeck coefficient changes significantly from positive to negative depending on the state of the Cβ-Cβ bonds.

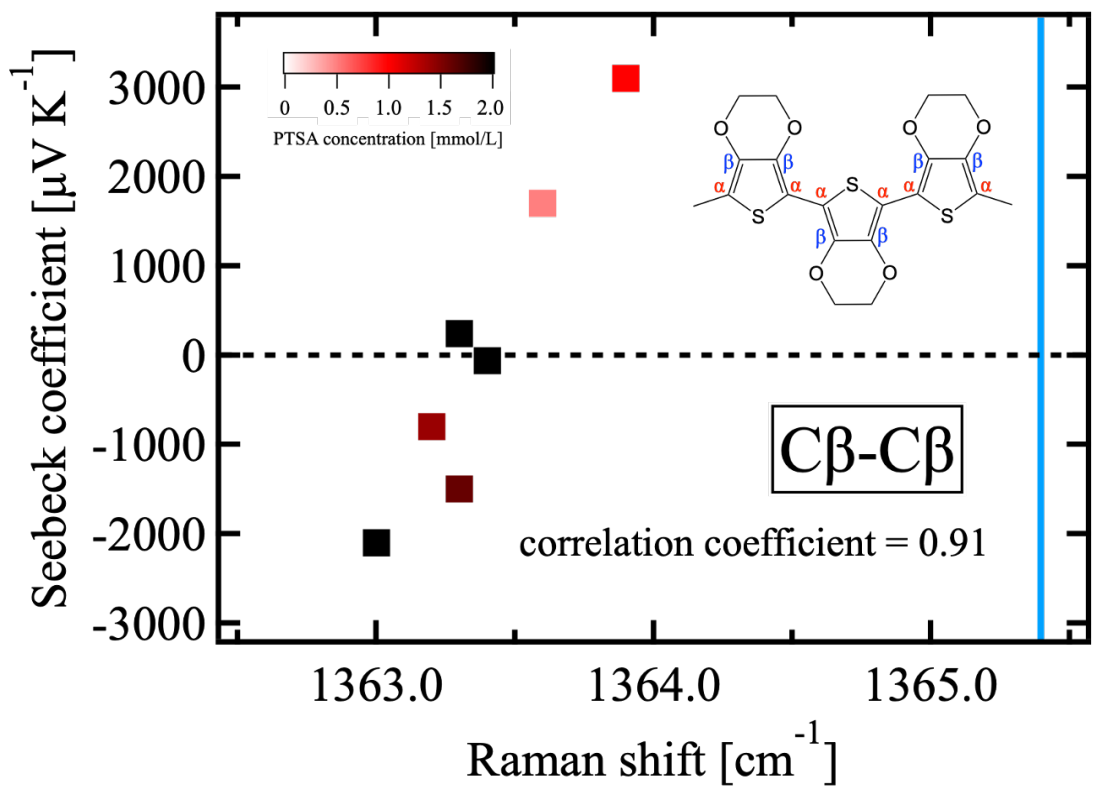

**Figure 7.** Seebeck coefficient of the PEDOT:PTSA sample on the felt fabric as a function of the Raman shift for the β-β bond. The solid blue line indicates the peak position of PEDOT:PSS/Glass. The sample variation comes from the PTSA concentration in the PEDOT-polymerization solution.

In fact, according to first-principles calculations, the DOS of PEDOT/PTSA has a sharp impurity level near the Fermi level, which the molecular distortion may further sharpen.[26]

## 3. Conclusion

The giant Seebeck effect, two orders of magnitude larger than that of conventional materials, was observed by loading PEDOT/PTSA onto a felt fabric made of PET fibers. The Raman scattering spectra show that the PEDOT molecules exhibit significant molecular distortion along the conduction path; specifically, the distortion of the Cβ-Cβ bond shows a strong correlation with the Seebeck coefficient. These results suggest that the density-of-states function of the PEDOT molecule was changed by molecular strain, and it is expected that the thermoelectric force of other conductive polymer materials can be increased based on the same principle.

**Acknowledgements**

This work was financially supported by Tateisi Science and Technology Foundation (No. 291904). The authors express sincere gratitude to Prof. Kanazawa (Tokyo City Univ.) for the fruitful discussion.